\documentclass[sigconf]{acmart}

\usepackage[russian, english]{babel}
\usepackage{xcolor}
\AtBeginDocument{%
  \providecommand\BibTeX{{%
    \normalfont B\kern-0.5em{\scshape i\kern-0.25em b}\kern-0.8em\TeX}}}

\setcopyright{acmcopyright}
\copyrightyear{2018}
\acmYear{2018}
\acmDOI{10.1145/1122445.1122456}

\begin{document}

\author{Vasyl Pihur}
\affiliation{%
  \institution{Snap, Inc}
  \country{USA}
}
\email{vpihur@snapchat.com}

\title{Query Processing at Snapchat: How we Handle Query Completion, Suggestion and Localization}


\begin{abstract}
Snapchat, a popular social media platform, started as a simple communication application. Over the years, however, it introduced Lenses and filters, Stories, Bitmoji, Stickers, Discover publishing content, Maps, Cameos and recently, Spotlight. With the ever increasing breadth of new engaging features came the need for an efficient and universal search platform. Learnings from a wider industry were quickly adopted, but a few challenges remained. With an average search query of just over 4 characters long, query completions, suggestions and localization left a lot of room for innovation. In this work, we present a Query Processing Layer (QPL), designed and implemented at Snap as part of our universal search platform.
\end{abstract}

\begin{CCSXML}
<ccs2012>
<concept>
<concept_id>10002951.10003317.10003325.10003328</concept_id>
<concept_desc>Information systems~Query log analysis</concept_desc>
<concept_significance>500</concept_significance>
</concept>
<concept>
<concept_id>10002951.10003317.10003325.10003329</concept_id>
<concept_desc>Information systems~Query suggestion</concept_desc>
<concept_significance>500</concept_significance>
</concept>
</ccs2012>
\end{CCSXML}

\ccsdesc[500]{Information systems~Query log analysis}
\ccsdesc[500]{Information systems~Query suggestion}

\keywords{query processing, spell-correction, localization, information retrieval}


\begin{teaserfigure}
  \includegraphics[width=\textwidth]{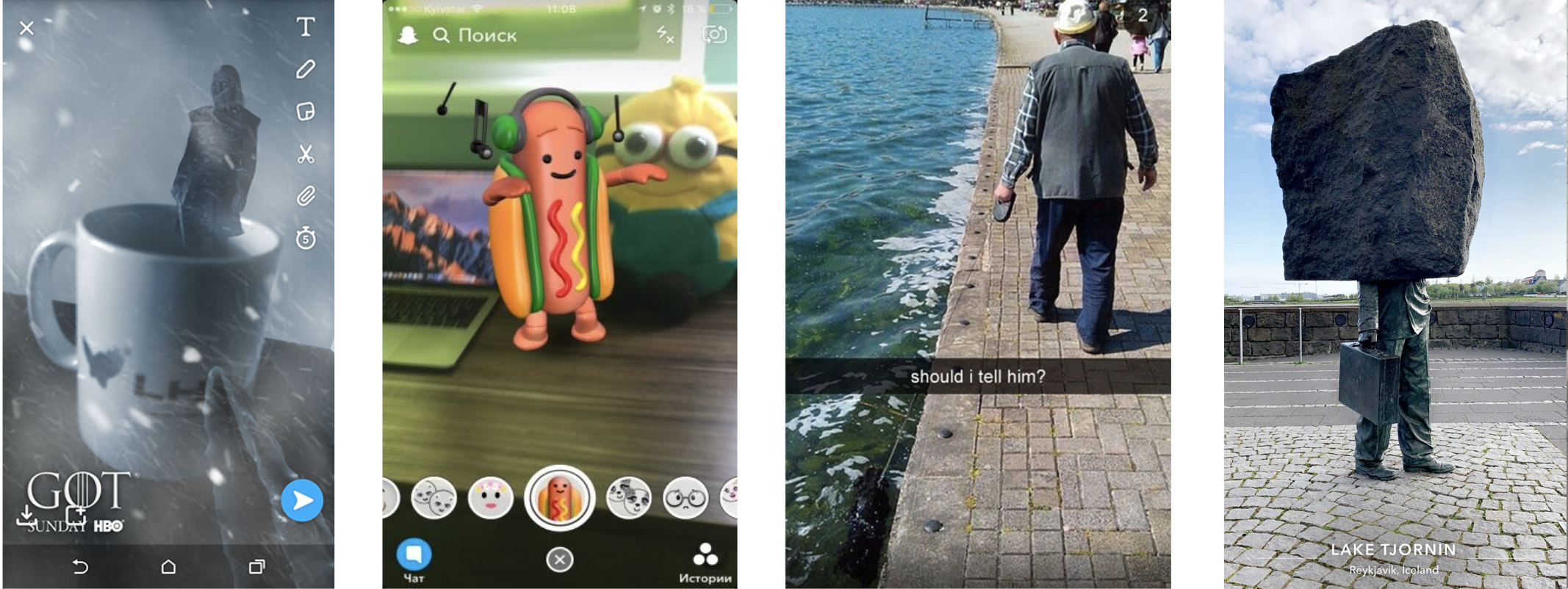}
  \Description{Examples of visual communication capabilities on the Snapchat platform.}
  \label{fig:teaser}
\end{teaserfigure}

\maketitle

\section{Introduction}

\begin{figure*}[!ht]
  \includegraphics[width=\textwidth]{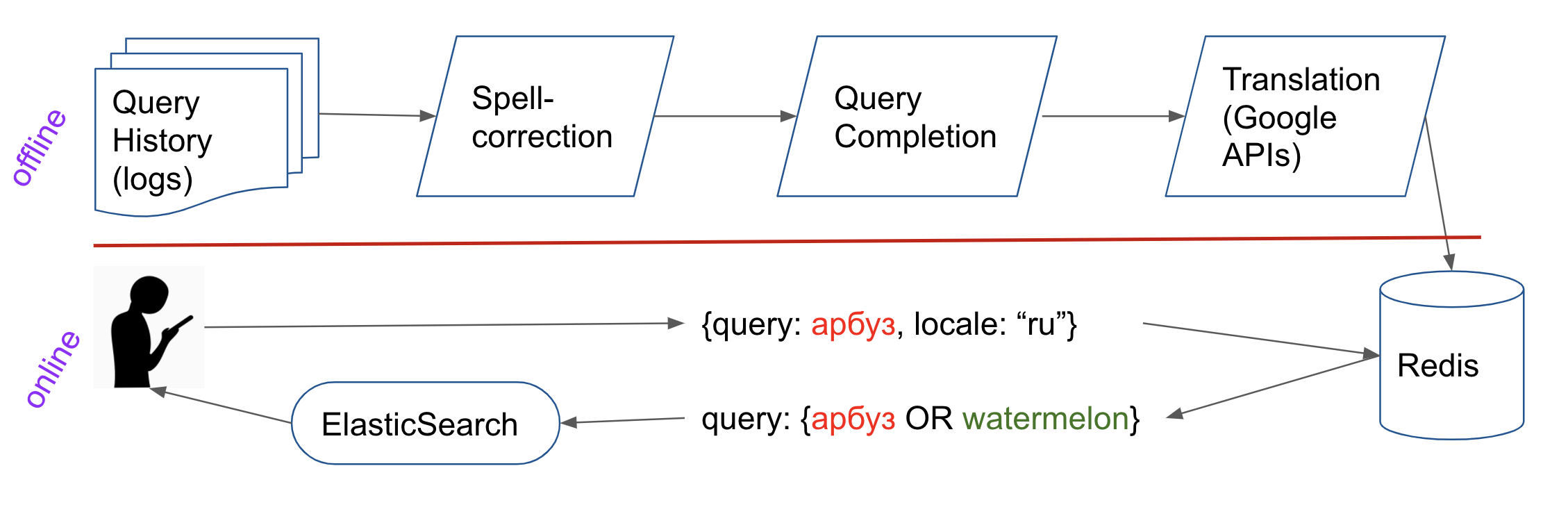}
  \caption{Offline and online components of QPL.}
  \label{fig:flow}
\end{figure*}

Snapchat prides itself on enabling its users to communicate visually, taking the adage saying ``a picture is worth a thousand words" to a whole new level. All of the new viral features that came out over the past several years reinforce this belief: Lenses, Stickers, Bitmoji, Cameos, Spotlight and several more. Searching for this visual content on the platform, however, remains a challenge. 

To help our users discover everything Snapchat has to offer, we take a two-pronged approach, similarly to just about any other company out there faced with the same problem. First, we attempt to anticipate what users would like to engage with by promoting popular and trending content\cite{Bawab2012FindingTL}, as well as using machine learning for finer personalized suggestions\cite{6648327, pers}. Second, we offer a search product that allows users to explicitly tell us what they want. We are in the very early stages of content understanding and enabling our users to search in novel and empowering ways (how can we make searching for snaps shown in the teaser figure above easy?) will be the focus of future work. At the present time, we rely mostly on manual content tagging, usually performed at the time of content creation, to generate a textual representation of a piece of visual document. The search engine matches user textual queries with content tags, but many challenges remain unaddressed in this simple setup.

The first major challenge is that Snapchat is a mobile-first platform, meaning that most users engage with the application on a reasonably small device screen. Typing on a mobile keyboard is tiresome and error-prone\cite{typo} and the frequency and variation in typing errors we see in the data is substantial. 

The second major challenge is the length of our queries. Users, on average, take just over 4 keystroke actions before making a selection. Traditional natural language processing (NLP) query understanding and complex semantic analyses\cite{nlp} yield little benefit under these conditions. 

The third major challenge is localization\cite{i18n}. Because of the visual nature of most of our content, it transcends the linguistic and social borders and is, for the most part, globally understood and appreciated. But someone searching in Spanish will not be able to find a dancing hotdog lens (second snap in the teaser figure) that is tagged in English, \emph{unless} its tagging keywords (``hotdog", ``dancing", etc) are explicitly translated into Spanish and included in the index. This process is expensive, time-consuming and inefficient.

And lastly, wouldn't it be amazing to \emph{visually} search for visual content? A lot of work remains to be done in this space, but our users already figured out that searching using emoji's\footnote{\href{https://en.wikipedia.org/wiki/Emoji}{https://en.wikipedia.org/wiki/Emoji}} (emoji keyboards are quite prevalent at this time) is convenient and takes only one character! One should be able to enter an emoji ``camel" character to get a set of camel lenses for us to consider that we made at least some progress on this front, which would be an incremental step towards the end goal.

The Query Processing Layer (QPL) was designed to address the challenges discussed above and enable the first iteration of visual search on the Snapchat platform. QPL was launched in February of 2021 for Lenses, Bitmoji and sticker search and is now part of our universal search platform available to all users.

In section \ref{sec:offline}, we will discuss how the offline component of the QPL is designed and implemented. Section \ref{sec:online} will describe the online querying aspect of the system. In Section \ref{sec:results}, we will present some results on how QPL is performing in practice and we will then finish with a discussion in Section \ref{sec:discussion}.

\section{How Query Processing Layer works?} 
Figure \ref{fig:flow} presents a visual diagram for both \emph{offline} and \emph{online} components of QPL, where everything begins and ends with users making their search requests. The offline component analyzes users' search behavior and aggregates its findings into a lookup table stored in redis\cite{redis}. The online component makes use of the data stored in redis to augment users' queries to achieve superior retrieval results\cite{ir}.

\subsection{The Offline Components of QPL}\label{sec:offline}
Our approach to query processing and enhancement\cite{qp1, qp2} is fully automated and data-driven and begins with understanding of common query patterns among our users. If a \emph{sufficiently large} number of users within each language cohort exhibits exactly the same search behavior, we aggregate their individual search events into query-level constructs\cite{segm}, keeping track of how these users arrived at their final target query.

\begin{table}
\centering
\begin{tabular}{ c }  
h \\
he \\
hes \\
he \\
hea \\
hear \\
heart \\
hear \\
hea \\
he \\
h \\
 \\
l \\
lo \\
lov \\
love \\
\end{tabular}
\caption{Hypothetical user session with two queries: ``heart" and ``love". We use a simple heuristics of looking for an empty query for keeping track of query boundary separation between the two intended queries.}
\label{tab:ex}
\end{table}

For example, Table \ref{tab:ex} shows a hypothetical search session with two queries where each row in the table represents a separate user action in a sequence, such as character addition or deletion. At this stage, we keep track of the mapping between ``subqueries'' and the target query, generating multiple string-to-string mappings \{``h" -> ``heart", ``he" -> ``heart", ``hes" -> ``heart", ``hea" -> ``heart", ``hear" -> ``heart", ``heart" -> ``heart", ``l" -> ``love", ``lo" -> ``love", ``lov" -> ``love", ``love" -> ``love"\}. We use a simple heuristic of the longest string to determine the target query, as well as remove duplicates resulting from character deletion.

Having processed all events, we compute an empirical estimate of conditional probabilities of all subquery to target query mappings, for example, ``hea" and ``heart", i.e. $P(``heart"|``hea")$. At this time, a threshold of 50\% is used to promote a candidate mapping into the next round. A careful reader will wonder why we promote mappings, such as ``love" -> ``love", which will become clear as we discuss the localization step.

\subsubsection{Spell correction}

\begin{figure}[!ht]
  \includegraphics[scale=0.28]{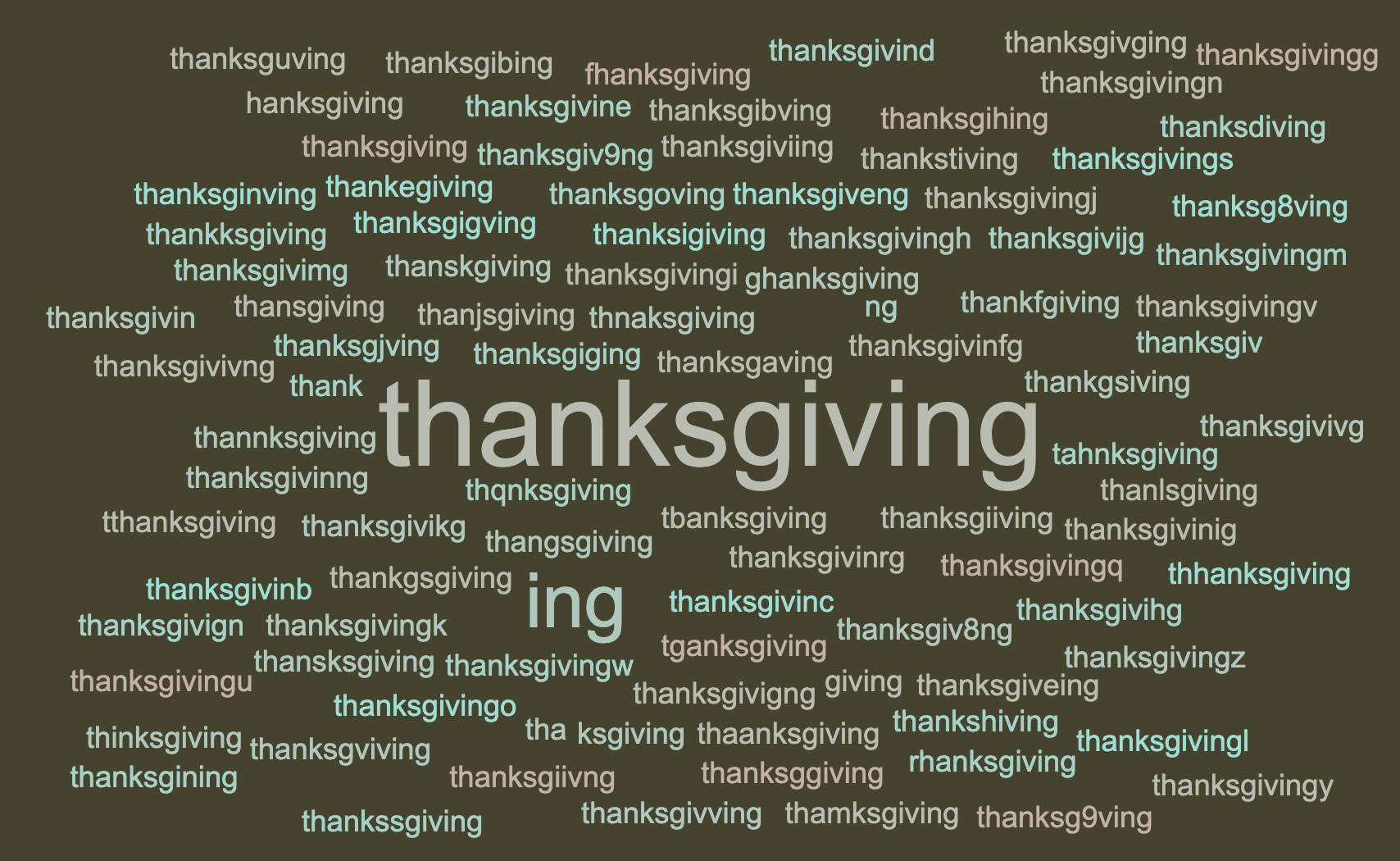}
  \caption{More than 100 common ways to misspell ``Thanksgiving". Users tend to accidentally hit the space bar instead of a character ``v" resulting in a high frequency of the ``ing" substring.}
  \Description{}
  \label{fig:thanks}
\end{figure}

We would like to keep common misspelling patterns\cite{spell} in the subquery space (such as ``hes" in Table \ref{tab:ex}), but our target queries ideally should match the tagging keywords in the index and, therefore, misspellings in those are highly undesirable.

Our users do not always adhere to the strict rules of any specific language and end up mixing languages, slang, shorthand or using outright creative ways of searching, such as emoji. One of our innovations in this work is our ability to perform spell-correction in a language-agnostic fashion, without using any outside reference dictionaries.

To perform spell corrections, we sort all target queries within each device language setting (locale) by their decreasing frequency of occurrence (most common query first). The very definition of a misspelling assumes that there must be the correctly spelled word with a higher frequency than the misspelling itself. After all, most users spell the word correctly! If there is no such word, then, by definition, the ``misspelling" in a very true sense is \emph{spelled} correctly. Taking this idea as a guiding principle, one proceeds by adding the most common word to an empty dictionary in proportion to its observed frequency. The second most common word is compared with the every word in the dictionary. If their distance\footnote{\href{https://norvig.com/spell-correct.html}{https://norvig.com/spell-correct.html}} is 1 and its frequency is 1\% of the first word, we consider the second word to be a misspelling of the first. It is replaced everywhere in the target mapping with the correctly spelled version and never added to the dictionary.

We then proceed to the next most common target query and so on. At each step, if the misspelling is identified, it is fixed in the target mapping, otherwise, it is added to the dictionary with the appropriate frequency count. The dictionary grows as the process is repeated for all queries and produces a language-specific, Snapchat-specific vocabulary that reflects the intended usage of the search platform. It contains ``good morning" and ``ttyl" as first class citizens, though none of these would be found in the English dictionary. Figure \ref{fig:thanks} shows over one hundred ``common" ways that our users misspell ``Thanksgiving", as an example of what kinds of misspelling patterns we observe.

\subsubsection{Query Completion}
We perform a simple aggregation\cite{autocomplete} step where we combine mappings that now point to the same target query due to changes in spell correction. For example,  ``thanksgivingg" -> ``tthanksgiving" and ``thanksgivingg" -> ``thanksgi ing" now should both point to the correctly spelled version and aggregated to remove duplicates.

\subsubsection{Localization}
The vast majority of our documents are visual. For example, a ``face swap" lens that replaces your face with another one or even an object, the linguistic and cultural barriers are negligible in terms of user being interested in them. But how does someone discover the ``face swap" lens using the Bahasa language in Indonesia? That lens would need to be tagged with appropriately translated equivalents in all languages where Snapchat is popular.

\begin{figure*}[!ht]
  \includegraphics[width=\textwidth]{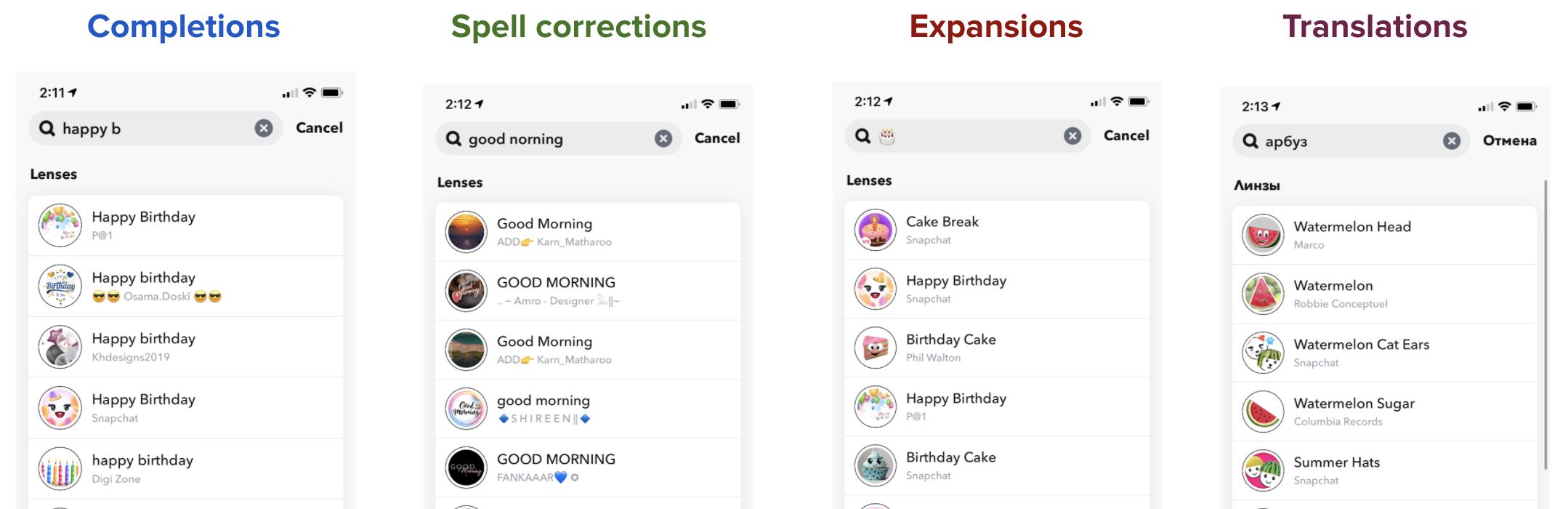}
  \caption{Four very different examples of QPL in action: a query completion for ``happy birthday", nuanced spell correction for ``good morning", emoji expansion and a translation to ``watermelon" from Russian.}
  \label{fig:inaction}
\end{figure*}

Instead of supporting full localization\cite{i18n} of every tagging keyword in English across all products and features, which is virtually impossible and very inefficient (after all most of these tags are likely to never meet an incoming query), we approach the problem from the other side. 

For all common target queries, we attempt a translation to English using Google Translate API's\footnote{\href{https://cloud.google.com/translate/docs/apis}{https://cloud.google.com/translate/docs/apis}}. If successful, Google returns an English equivalent, as well as detected language for the query. Correctly spelled queries significantly increase the likelihood of a successful translation which is why this step is performed last. It also becomes clear why we initially kept the ``A" -> ``A" mappings, since the second ``A" can now be easily replaced with its English equivalent.

\subsection{The Online Components of QPL} \label{sec:online}
In Figure \ref{fig:flow}, the bottom section demonstrates the online components of QPL, with the redis cluster shown on the very right. When the user issues a query, in this case, a Russian query for ``watermelon", a redis key is constructed as ``ru:\foreignlanguage{russian}{арбуз}" (notice how it is locale specific) and queried in redis, where a ``watermelon" is returned. The final user query is constructed with the original and QPL query combined using the OR operator. This query hits the ElasticSearch\cite{es} and the final set of results are sent to the user's device (as seen in the last panel in Figure \ref{fig:inaction}).

\section{QPL implementation and results at Snap} \label{sec:results}
The current size of our QPL map is anywhere between 4.5M and 5.5M entities and is rebuilt daily to accommodate changing tastes and preferences. Different languages, of course, have different coverage, depending on their levels of engagement. One of the key success metrics for us is to see significant growth in native searches in key markets.

Query completions comprise the largest portion of the QPL corpus and save users some tiresome typing by getting them to the desired results faster. The first panel of Figure \ref{fig:inaction} shows a query completion for ``happy birthday". Interestingly, the confidence of a completion varies by user's device language and we are able to make a suggestion much earlier in some locales versus others. Another advantage of the completion behavior is the newly-found stability of search results. We are now able to show almost the same results for ``bitm", ``bitmo", ``bitmoj", ``bitmoji", as all four queries get completed and dominated by the final ``bitmoji" keyword. 

Fixing typing errors is where QPL shines. The second panel in Figure \ref{fig:inaction} shows a spell correction of ``good norning" where a user hits an adjacent ``n" key instead of ``m". Prior to QPL, we showed no results for this very commonly misspelled search term.

The first seeds of visual search are implemented through emoji expansions where we map each emoji through the demoji Python package\footnote{\href{https://pypi.org/project/demoji/}{https://pypi.org/project/demoji/}} to its corresponding textual form. For example, notice how the birthday cake emoji in the third panel of Figure \ref{fig:inaction} gets expanded to ``birthday cake" and relevant results are displayed. Prior to this expansion, only lens names with the cake emoji in their names would match this query.

And lastly, localization has improved by orders of magnitude due to QPL, without significant investment in manual translations of tagging keywords. For some languages, such as Russian, French, Spanish, German, Korean, we saw double digits increases in content engagement after launching QPL. Smaller languages are likely to sustain similar gains, but it is hard to get a reliable measurement on a relatively small set of queries coming from users searching in them. 

At this time, QPL impacts anywhere between 15\% and 30\% of all queries, depending on the specific search product. With the small cost of a lookup (redis p95 latency is 1ms) and a relatively high hit rate, QPL has made a very large difference to search quality at Snap.

\section{Discussion} \label{sec:discussion}
Query processing at Snap takes place in a data-driven, adaptive and global fashion. We implemented QPL to be as efficient as possible, while preserving user's privacy\cite{rappor} by excluding unique and sensitive searches. For commonly made search requests, we utilize user data to give back to the community by saving users time and frustration. Cost effectiveness was also one of our key priorities when designing the system and we believe we have achieved that with the current system design: the backend runs on a single machine and redis is cheap and fast.

One of the most interesting characteristics of QPL is its self-evolving nature. The more QPL helps users discover content with ease, the more likely it becomes that users will increase their search engagement which, in turn, enriches the QPL mapping and expands its size and scope. This self-feedback loop almost guarantees that QPL will have a life of its own and will remain current and relevant by smoothly adopting to seasonal and trending changes.

We have many plans for the future of QPL. One of the most interesting directions is to expand the scope of QPL to more distant expansions and concept linkage. We are likely to dive deeper into generating the appropriate embedding space\cite{emb}, where the most relevant concept relationships can be discovered and linked based on their co-occurrence in user search sessions.

%\begin{acks}
%To Robert, for the bagels and explaining CMYK and color spaces.
%\end{acks}

\bibliographystyle{ACM-Reference-Format}
\bibliography{sample-base}

% \appendix
\end{document}